\pdfoutput=1

\documentclass[twocolumn,showpacs]{revtex4}
\usepackage{graphicx}
\usepackage{amssymb,amsmath}
\begin{document}

\title{
  Aharonov-Bohm and relativistic Corbino effects in graphene: A comparative study of two quantum interference phenomena
}

\author{Adam Rycerz\footnote{Corresponding author, 
    e-mail: \url{rycerz@th.if.uj.edu.pl}.}}
\affiliation{Instytut Fizyki im.\ Mariana Smoluchowskiego, 
Uniwersytet Jagiello\'{n}ski, Reymonta 4, PL--30059 Krak$\acute{o}$w, Poland}

\begin{abstract}
This is an analytical study of magnetic fields effects on the conductance, the shot noise power, and the third charge-transfer cumulant for Aharonov-Bohm rings and Corbino disks in graphene. The two distinct physical mechanisms lead to very similar magnetotransport behaviors. Differences are unveiled when discussing the third-cumulant dependence on magnetic fields.
\end{abstract}

\date{\today}
\pacs{ 72.80.Vp, 73.43.Qt, 73.63.-b }
\maketitle

{\it Introduction.}---The advent of graphene, two-dimensional form of carbon in which itinerant electrons behave as massless Dirac fermions \cite{Sem84,Nov05}, led condensed-matter physicists to reexamine effects of quantum transport in nanostructures \cite{Naz09}. The Aharonov-Bohm effect (ABE) \cite{Imr89}, a famous condensed-matter realization of the two-slit gedankenexperiment \cite{Pop02}, has also gained some attention \cite{Rec07,Rus08,Ryc09}. Very recently, it was predicted theoretically \cite{Ryc10} that periodic (approximately sinusoidal) magnetoconductance oscillations appear in weakly doped Corbino disks in graphene. Unlike ABE, which also appears for Schr\"{o}dinger electrons in the two-dimensional electron gas (2DEG), the {\em quantum relativistic Corbino effect} (QRCE) is a quantum-interference phenomena specific for massless Dirac fermions, for which transmission via evanescent waves leads to a finite value of the conductance at zero doping. 

\begin{figure}
\centerline{\includegraphics[width=0.9\linewidth]{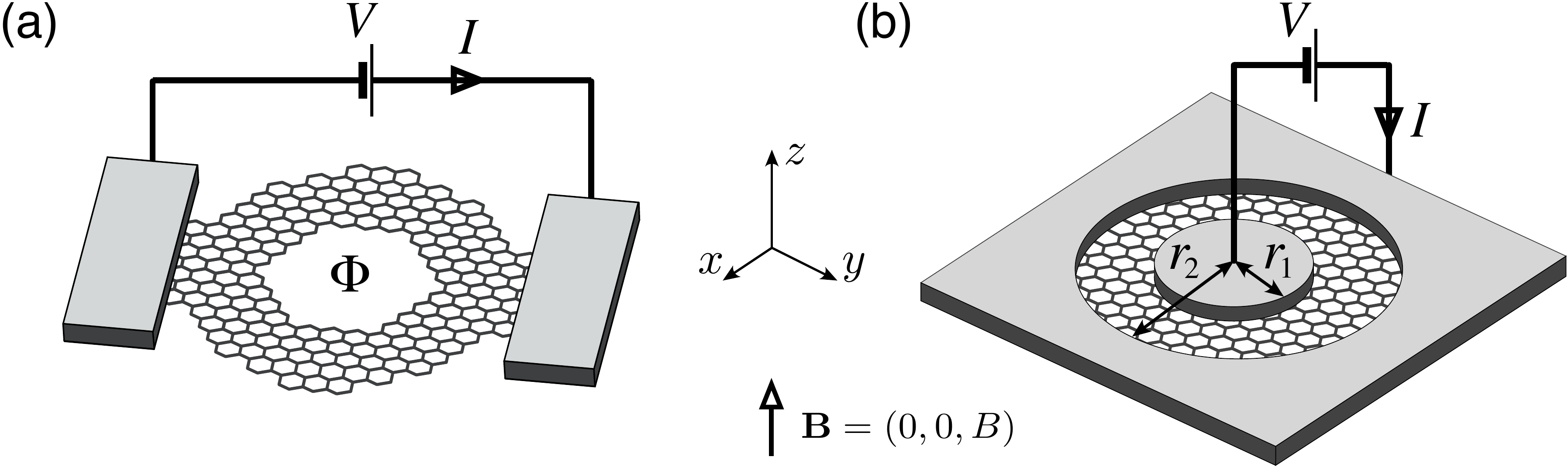}}
\caption{ 
  Devices considered in the paper (schematic). Voltage source passes the current from to the right to the left lead in case of the Aharonov-Bohm ring (a) or from the outer circular lead to the inner one in case of the Corbino disk (b) in graphene. The coordinate system and the applied magnetic field orientation (same for both devices) are also depicted. Additional gate electrodes (not shown) may be used to tune dopings or to induce transverse electric fields. 
  \label{fig:setup}}
\end{figure}

It was also found in Refs.\ \cite{Ryc10} and \cite{Kat10} that for QRCE in disks of moderate radii ratios $r_2/r_1\lesssim{10}$ (see Fig.\ \ref{fig:setup}) two basic transport characteristics: the conductance $G$ and the shot-noise power (quantified by the Fano factor $F$) show qualitatively similar behavior as their ABE counterparts. In this paper we extent the discussion on the third charge-transfer cumulant \cite{Reu03} showing that this quantity (analyzed as a function of applied magnetic field) demonstrates several new features of QRCE absent in ABE, including the oscillations frequency doubling at $r_2/r_1\simeq{7}$. But first, we briefly recall the basic definitions of mesoscopic electron transport characteristics.

{\it Counting statistics and charge cumulants.}---Electric charge $Q$ passing the nanoscale device at the time interval $\Delta{t}$ in the presence of external bias voltage $V$ is a random variable, a statistical distribution of which can be expressed in terms of characteristic function
\begin{equation}
  \Lambda(\chi)=\left<\,\exp(i\chi{Q}/e)\,\right>,
\end{equation}
where the electron charge is $-e$  and $\langle{X}\rangle$ denotes the expectation value of $X$. In the so-called {\em shot-noise} limit $eV\gg{k_BT}$ (with Boltzman constant $k_B$ and temperature $T$) the characteristic function is given by \cite{Naz09}
\begin{equation}
  \ln\Lambda(\chi)=\frac{seV\Delta{t}}{2\pi\hbar}\sum_p
  \ln\left[1+T_p\left(e^{i\chi}-1\right)\right],
\end{equation}
where $T_p$ are transmission probabilities for normal modes in leads, each of which has a degeneracy $s$ (at low fields, $s=2$ for 2DEG, or $s=4$ for bulk graphene due to spin and valley degeneracy). We have assumed $V>0$ for simplicity; $\hbar$ denotes the Plank constant. 

The average charge $\langle{Q}\rangle$, as well as higher charge-cumulants $\langle\langle{Q^k}\rangle\rangle\equiv\langle\,(Q-\langle{Q}\rangle)^k\,\rangle$ may be obtained by subsequent differentiation of $\ln\Lambda(\chi)$ with respect to $i\chi$ and setting $\chi=0$. In particular, the conductance
\begin{equation}
  \label{gdef}
  G\equiv\frac{\langle{Q}\rangle}{V\Delta{t}}=\frac{e}{V\Delta{t}}
  \left.\frac{\partial\ln\Lambda}{\partial(i\chi)}\right|_{\chi=0}=
  \frac{se^2}{2\pi\hbar}\sum_pT_p,
\end{equation}
which restores the Landauer-Buttiker formula. Analogously, the Fano factor
\begin{equation}
  \label{fdef}
  F\equiv\frac{\langle\langle{Q^2}\rangle\rangle}{\langle\langle{Q^2}\rangle\rangle_{\rm Poisson}}=\frac{\sum_pT_p(1-T_p)}{\sum_pT_p},
\end{equation}
where $\langle\langle{Q^2}\rangle\rangle_{\rm Poisson}=e\langle{Q}\rangle$ denotes the value of $\langle\langle{Q^2}\rangle\rangle$ for the Poissonian limit $T_p\ll{1}$. We further define the $R$-factor, quantifying the third cumulant 
\begin{equation}
  \label{rdef}
  R\equiv\frac{\langle\langle{Q^3}\rangle\rangle}{\langle\langle{Q^3}\rangle\rangle_{\rm Poisson}}=\frac{\sum_pT_p(1-T_p)(1-2T_p)}{\sum_pT_p},
\end{equation}
with $\langle\langle{Q^3}\rangle\rangle_{\rm Poisson}=e^2\langle{Q}\rangle$. For undoped graphene samples, similarly as for diffusive wires, the distribution of transmission eigenvalues is $\rho_{\rm diff}(T)=2G/(g_0T\sqrt{1-T})$ \cite{Bee08}, with the conductance quantum $g_0\equiv{2e^2/\pi\hbar}$, leading to $F=1/3$ and $R=1/15$.

\begin{figure}
\centerline{\includegraphics[width=0.7\linewidth]{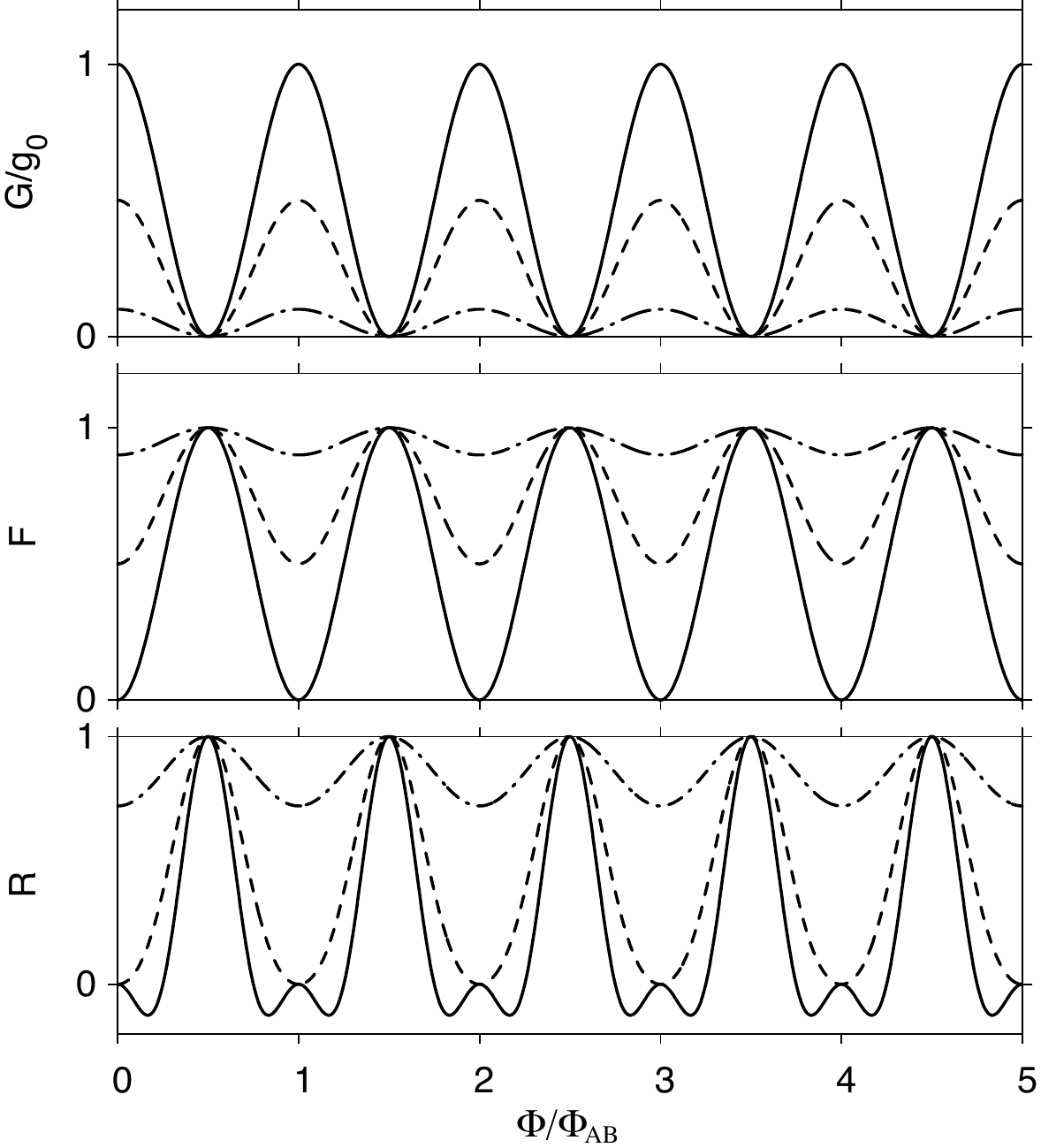}}
\caption{
  Magnetic flux effect on the conductance (top panel), the Fano factor (middle panel) and the $R$-factor (bottom panel) obtained from a single-mode model of transport via symmetric Aharonov-Bohm ring (\ref{tptunab}). The parameters are: $\Gamma=1$ (solid lined), $\Gamma=1/2$ (dashes lines), and $\Gamma=1/10$ (dashed-dot lines); $\gamma_0=0$ for all curves. 
\label{fig:abri}}
\end{figure}

{\it Aharonov-Bohm effect in graphene.}---Recent experimental \cite{Rus08} and numerical \cite{Ryc09} studies suggest that the magnetoconductance of Aharonov-Bohm rings in graphene behaves similarly as the magnetoconductance of two parallel tunnel junctions \cite{Naz93}. In particular, the oscillations magnitude $\Delta{G}\equiv{G_{\rm max}-G_{\rm min}}\propto{G}_{\rm av}\ll{g_0}$ (with $G_{\rm max}$, $G_{\rm min}$, and $G_{\rm av}$ the maximal, the minimal and the average value of the conductance when varying flux piercing the ring area $\Phi$; see Fig.\ \ref{fig:setup}(a)). 

Such observations allow us to regard electron transport through a narrow symmetric ring as dominated by a single mode ($p=0$) with
\begin{equation}
  \label{tptunab}
  T_0=\Gamma\cos^2\left(\frac{\gamma_0}{2}+\frac{\pi\Phi}{\Phi_{\rm AB}}\right),
\end{equation}
where $\Gamma\leqslant{}1$ is the transmission probability for each of the ring arms \cite{gamfoo}, $\gamma_0$ is the total dynamic phase gained by an electron traveling around the ring at zero magnetic field (typically, $\gamma_0$ is controlled by the transverse electric field induced be gate electrodes not shown in Fig.\ \ref{fig:setup}(a)), and $\Phi_{\rm AB}\equiv{2\pi\hbar/e}$ is the flux quantum. We notice here, that randomly-chosen $\gamma_0$ leads to the distribution of transmission probabilities $\rho_{\rm AB}(T)=1/\pi\sqrt{T(\Gamma-T)}$, with $0\leqslant{T}\leqslant{\Gamma}$. Subsequently, $G=g_0\Gamma/2$, $F=1/4$, and $R=0$, which reproduce the results for a symmetric chaotic cavity \cite{Naz09}. Although we consider a simple model (\ref{tptunab}), the universality of chaotic cavity transport properties lets us believe, that these results also hold true for real systems. So far, the experimental values of $F$ and $R$ for Aharonov-Bohm rings in graphene are unavailable.

Substituting $T_p=T_0\delta_{0,p}$ into Eqs.\ (\ref{gdef}--\ref{rdef}) we obtain the conductance, the Fano factor, and the $R$-factor presented in Fig.\ \ref{fig:abri} for $\gamma_0=0$ and selected values of $\Gamma$. Each of the studied quantities oscillates as a function of flux with period $\Phi_{\rm AB}$. The influence of higher harmonic frequencies become visible only for the $R$-factor if $\Gamma\gtrsim{1/2}$. The first-harmonic amplitude, however, dominates the magnetic field dependence of $R$ for any $\Gamma$, and no qualitative effect (such as the frequency doubling) is present. Also, the oscillations magnitudes: $\Delta{G}/g_0=\Delta{F}=\Gamma$ and $\Delta{R}=\Gamma'(3-2\Gamma')$, with $\Gamma'=\min(\Gamma,\frac{3}{4})$, are all monotonically increasing functions of $\Gamma$.

\begin{figure}
\centerline{\includegraphics[width=0.7\linewidth]{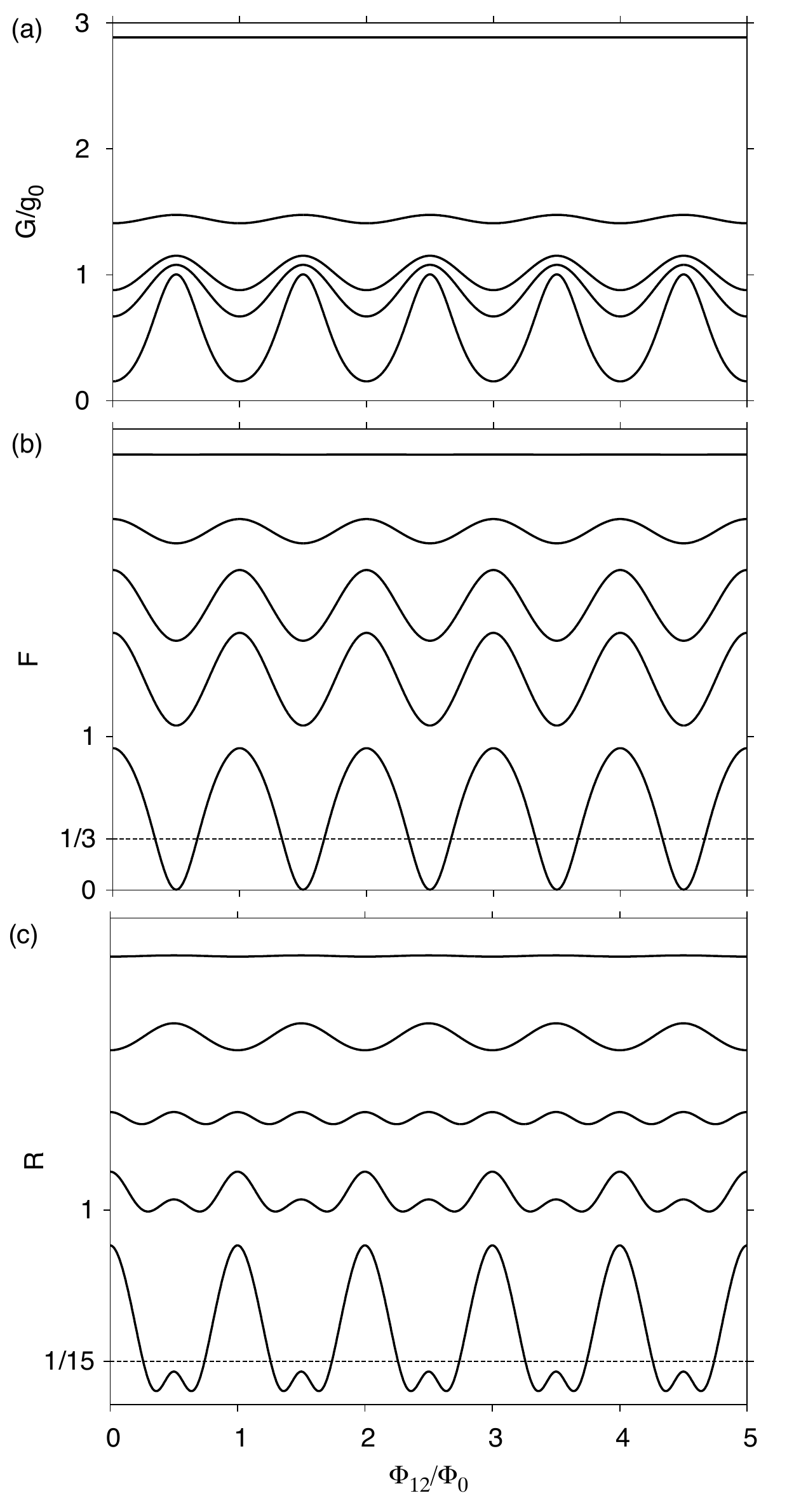}}
\caption{
  Same as in Fig.\ \ref{fig:abri} but for the undoped Corbino disks of radii ratios $r_2/r_1=2$, $4$, $7.2$, $10$, and $50$ (top to bottom solid line at each panel). On panels (b) and (c) all but the uttermost bottom curves are shifted upward for clarity. Diffusive values of $F$ and $R$ are also shown (dashed lines).
\label{fig:corbi}}
\end{figure}

{\it Relativistic Corbino effect.}---Transmission probabilities for the Corbino disk in undoped graphene (see Fig.\ \ref{fig:setup}(b)) are given by \cite{Ryc10}
\begin{equation}
\label{tjcorb}
  T_j=\frac{1}{\cosh^2\left[(j+\Phi_{12}/\Phi_0)\ln(r_2/r_1)\right]},
\end{equation}
where $j=\pm\frac{1}{2},\pm\frac{3}{2},\dots$ is the angular momentum quantum number, $\Phi_{12}=\pi{B}(r_2^2-r_1^2)$ is the flux piercing the disk in the uniform magnetic field $B$ \cite{mfoot}, and $\Phi_0\equiv{2}\Phi_{\rm AB}\ln(r_2/r_1)$ is the oscillation period of the conductance and higher charge-cumulants. Eq.\ (\ref{tjcorb}) holds true also in the small vicinity of the charge-neutrality point, defined via $|\Phi_{12}|\lesssim{2\Phi_{\rm AB}}\ln(\hbar{v_F}/|\mu|r_1)$, where $v_F\simeq{10^6}\,$m/s is the Fermi velocity in graphene and $\mu$ is the electrochemical potential.

We now substitute $T_j$ to Eqs.\ (\ref{gdef}--\ref{rdef}) for measurable quantities. Results are presented in Fig.\ \ref{fig:corbi}. For small values of $r_2/r_1$ we have $G\simeq{2g_0}/\ln(r_2/r_1)$, $F\simeq{1/3}$, and $R\simeq{1/15}$, reproducing the diffusive values. This is also worth to notice, that taking random fluxes $\Phi_{12}$ covering uniformly the period $\Phi_0$, one restores $\rho_{\rm diff}(T)$ for arbitrary $j$ and $r_2/r_1$ in Eq.\ (\ref{tjcorb}). Subsequently, diffusive values of $G$ and higher charge cumulants $\langle\langle{Q^k}\rangle\rangle$ are exactly restored by averaging over magnetic fields for any $r_2/r_1$. For large $r_2/r_1$, the system alternates between two transport regimes when varying magnetic field: At $\Phi_{12}/\Phi_0$ close to half-odd integer, the transport is govern by a mode with $T_j\simeq{1}$, leading to $G\simeq{g_0}$,  $F\simeq{R}\simeq{0}$, such as for a single-mode quantum point contact. On the contrary, when  $\Phi_{12}/\Phi_0$ is close to an integer, we have $T_j\ll{1}$ for all $j$-s, leading to $G\ll{g_0}$, $F\simeq{R}\simeq{1}$, such as for a tunneling junction. This feature makes the system behavior similar to that characteristic for a double barrier rather than the Aharonov-Bohm ring \cite{Naz09}.

\begin{figure}
\centerline{\includegraphics[width=0.7\linewidth]{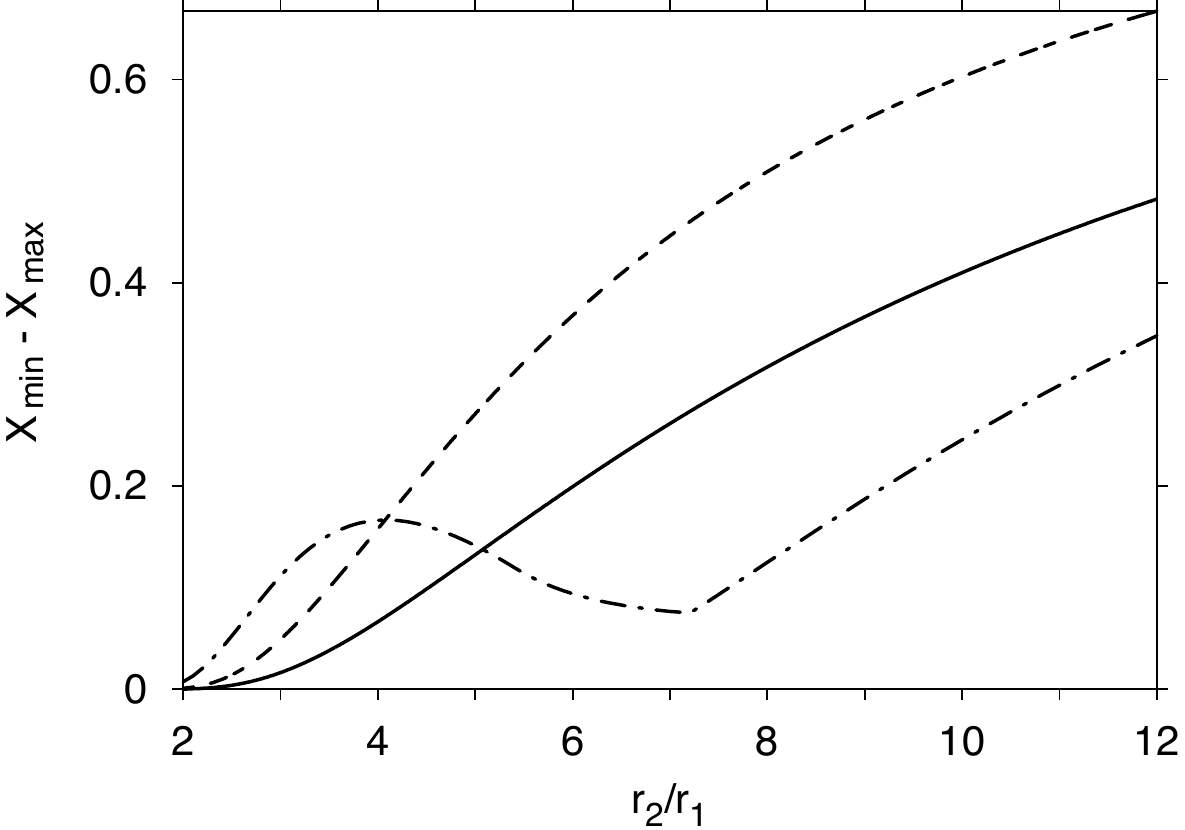}}
\caption{
  Oscillations magnitude $X_{\rm max}\!-\!X_{\rm min}$ for $X=G/g_0$ (solid line), $X=F$ (dashed line), and $X=R$ (dashed-dot line) as a function of the Corbino disk radii ratio $r_2/r_1$.
\label{fig:del}}
\end{figure}

{\it Fourier analysis.}---A deeper insight into the magnetic fields effect on transport via Corbino setup is provided with Fourier decomposition of the measurable quantities. The following expressions are obtained employing the Poisson theorem:
\begin{align}
G &= G_0\left(1+2\sum_{n\geq{1}}\nolimits\alpha_n\cos\phi_n\right), \label{gserie} \\
F &= \frac{\frac{1}{3}+2\sum_{n\geq{1}}\alpha_n\left(\frac{1}{3}-\frac{2\pi^2}{3}l_n^2\right)\cos\phi_n}{1+2\sum_{n\geq{1}}\alpha_n\cos\phi_n}, \label{fserie} \\
R &= \frac{\frac{1}{15}+2\sum_{n\geq{1}}\alpha_n\left(\frac{1}{15}-\frac{2\pi^2}{3}l_n^2+\frac{4\pi^4}{15}l_n^4\right)\cos\phi_n}{1+2\sum_{n\geq{1}}\alpha_n\cos\phi_n}, \label{rserie} 
\end{align}
with $G_0\equiv{2g_0}/\ln(r_2/r_1)$, $\phi_n\equiv{2}\pi{n}\Phi_{12}/\Phi_0$, $l_n\equiv{n}/\ln(r_2/r_1)$, and $\alpha_n\equiv{(-1)^n}\pi^2l_n/\sinh(\pi^2l_n)$.

It is clear from Eqs.\ (\ref{gserie}--\ref{rserie}), that for the moderate values of $r_2/r_1$ (corresponding to large $l_n$-s) the oscillations magnitude of first three charge-cumulants grows systematically with the cumulant rank (see Fig.\ \ref{fig:del}). In fact, $\Delta{X}\equiv{X_{\rm max}-X_{\rm min}}$ exceed $10\%$ of the diffusive value for $r_2/r_1>{4.9}$ if $X=G/g_0$, for $r_2/r_1>{2.8}$ if $X=F$, and for $r_2/r_1>{2.0}$ if $X=R$. The behavior of charge cumulants for larger $r_2/r_1$ is, however, more complicated. Expanding Eqs.\ (\ref{gserie}--\ref{rserie}) up to the terms $\propto{|\alpha_1|}$ gives us 
\begin{align}
  G &\simeq G_0\left[1-2|\alpha_1|\cos\left(2\pi\Phi_{12}/\Phi_0\right)\right],  \\
  F &\simeq \frac{1}{3} + \frac{4\pi^2}{3}|\alpha_1|\,l_1^2\cos\left(2\pi\Phi_{12}/\Phi_0\right),  \\
  R &\simeq \frac{1}{15} + \frac{4\pi^2}{3}|\alpha_1|\left(l_1^2-\frac{2\pi^2}{5}l_1^4\right)\cos\left(2\pi\frac{\Phi_{12}}{\Phi_0}\right).
\end{align}
Due to a fast decay of $|\alpha_n|$ with growing $n$, the above provides excellent approximations of $G$ and $F$ for $r_2/r_1\lesssim{10}$. This is not the case for $R$, a first-harmonics amplitude of which changes the sign at $r_2/r_1=\exp(\sqrt{2/5}\pi)\simeq{7.2}$, leading to the oscillations frequency doubling (see Fig.\ \ref{fig:corbi}(c)) and to a cusp-shaped local minimum on the $\Delta{R}$ plot (dashed-dot line in Fig.\ \ref{fig:del}, notice also a local maximum at $r_2/r_1\simeq{4}$). For larger $r_2/r_1$-s, the magnitudes grows slowly approach the limiting value $\Delta{G}/g_0=\Delta{F}=\Delta{R}=1$, keeping the relation $\Delta{R}<\Delta{G}/g_0<\Delta{F}$.

{\it Conclusions.}---We find the third cumulant of charge transfer via the Corbino magnetometer in graphene with moderate (and thus most likely experimentally accessible) outer to inner radii ratios $r_2/r_1$ exhibits remarkably stronger oscillations with the varying field then earlier predicted for the conductance \cite{Ryc10} and the shot-noise power \cite{Kat10}. The oscillations magnitude shows surprising features when studied as a function of $r_2/r_1$, including the maximum at $r_2/r_1\simeq{4}$,  and cusp-shaped minimum accompanied by frequency doubling near $r_2/r_1\simeq{7}$. Such features, together with the size-dependent oscillations period, may help determining the effective proportions of ballistic graphene samples attached to metallic leads, at least in the case of a rotationally-symmetric setup.

Additionally, third cumulant appears to be the lowest in rank that is capable of illustrating the qualitative difference between electron magnetotransport through the Aharonov-Bohm and the Corbino quantum interference devices in graphene.
Certain features of such two systems suggest that the former and the latter may reproduce transport properties of chaotic cavities and diffusive wires (respectively).

{\it Acknowledgments.}---The work was supported by the National Science Centre of Poland (NCN) via Grant No.\ N--N202--031440, by the Alexander von Humboldt Stiftung-Foundation, and partly by Foundation for Polish Science (FNP) under the program TEAM.


\begin{thebibliography}{}
\bibitem{gamfoo}
Typically, $\Gamma\ll{}1$ as for graphene rings with irregular edges there are only evenescent modes present in each of the ring arms at zero and weak dopings, see Refs.\ \cite{Rus08,Ryc09}.
\bibitem{mfoot}
More generally, for centrosymmetric field ${\bf B}=B(r)\hat{e}_z$, the parameter $\phi={\Phi_{12}/\Phi_0}$ in Eq.\ (\ref{tjcorb}) is replaced by \cite{Kat10}
$$
  \phi'\equiv\frac{\Phi_1}{\Phi_{\rm AB}}+\frac{4\pi}{\Phi_0}\int_{r_1}^{r_2}
  \frac{dr}{r}\int_{r_1}^r{dr'}r'B(r),
$$
where $\Phi_1$ is flux through the inner ring ($r<r_1$). In particular, if $B(r)=0$ for $r>r_1$, the Aharonov-Bohm-like oscillations occur when varying $\Phi_1$.
\end{thebibliography}
\end{document}